\documentclass[prd,nofootinbib,showpacs,preprint,aps]{revtex4-1}
\usepackage{subfigure}
\usepackage{amsmath}
\usepackage{amsfonts}
\usepackage{amssymb}
\usepackage[dvips]{graphicx}
\usepackage{advdate}

\usepackage[applemac]{inputenc}
\usepackage[T1]{fontenc}
\usepackage{amssymb}
\usepackage{amsfonts}
\usepackage{amsfonts} 
\usepackage{graphicx} 

\usepackage{bbm}
\def\beq{\begin{equation}}
\def\eeq{\end{equation}}
\def\bsp{\begin{split}}
\def\esp{\end{split}}
\def\bea{\begin{eqnarray}}
\def\eea{\end{eqnarray}}
\def\ba{\begin{array}}
\def\ea{\end{array}}
\def\nn{\nonumber \\}

\def\lb{\left(}
\def\rb{\right)}

\def\l.{\left.}
\def\r.{\right.}

\def\la{\langle}

\def\part{\partial}

\def\m{{\mu}}
\def\n{{\nu}}

\def\ie{{\it i.e. }}
\def\ame{&=&}
\def\nn{\\ \nonumber}

\begin{document}
\preprint{UdeM-GPP-TH-25-305}
\title{Graviton Scattering on Gravitational Atoms: Relic Graviton Shot Noise}
\author{Benjamin Avila-Lopez $^{1,2}$}
\email{benjamin.avila@usach.cl$^1$}
\author{Richard MacKenzie$^{2,3}$}
\email{richard.mackenzie@umontreal.ca}
\author{Fernando Mendez$^1$}
\email{feritox@gmail.com}
\author{M. B. Paranjape$^{2,3,4,5}$} 
\email{paranj@lps.umontreal.ca}
\affiliation{$^{1}$Departamento de F\'\i sica, Universidad de Santiago de Chile, Avenida Ecuador 3493, Estacion Central, Santiago, Chile 9170124}
\affiliation{$^{2}$Groupe de physique des particules, D\'epartement de physique,$^3$Institut Courtois,$^4$Centre de recherche math\'ematiques,  Universit\'e de Montr\'eal, C.P. 6128, succ. centre-ville, Montr\'eal, Qu\'ebec, Canada, H3C 3J7 }
\affiliation{$^{5}$Department of Physics, University of Auckland, Auckland, New Zealand}
\begin{abstract}\baselineskip=18pt
We study the scattering of a graviton on a gravitational atom.  By gravitational atom we mean a quantum mechanical system of a gravitational (bound) state of two massive particles, with possibly some boundary conditions (such as bouncing on a table or hanging as a pendulum).    We demonstrate the unexpected fact, that the total absorption cross section is universal, it is independent of both the mass of the gravitationally bound particle or of the mass providing the binding potential.  We find that the total absorption cross-section is simply  proportional to the Planck area, multiplied by a dimensionless, numerical factor.  We speculate about the potential for detection of relic gravitons shot noise.

\vskip 1in
\leftline{Essay written for the Gravity Research Foundation 2025 Awards for Essays on Gravitation.}
\leftline{March 31st, 2025}
\leftline{Corresponding author: M. B. Paranjape, paranj@lps.umontreal.ca}
\end{abstract}
\maketitle
\section{Introduction}     
Gravitons scattering on quantum mechanical systems can induce transitions between the quantum levels of the system, corresponding to absorption or stimulated emission of gravitons.  When the quantum mechanical system is infact gravitational, we find the result that the total absorption or spontaneous emission cross section is independent of the masses of the elements that make up the quantum mechanical system.  

We first discovered this result for the case of gravitons interacting with ultra-cold neutrons that were quantum mechanically bouncing on the surface of a table \cite{gl}.  The cross section must have units of area, and as there are no other dimensionful parameters in the model, the only available area is the Planck area, $\frac{\hbar G}{c^3}$.  Indeed, we found that the cross section was proportional to the Planck area multiplied by a purely numerical factor.  

In this essay  we show the analogous result for gravitons scattering on a gravitational atom.  A gravitational atom corresponds to the system where one massive particle is quantum mechanically, gravitationally, bound to another massive particle.  We do the calculation using Newton's law for the gravitational potential, however, this must be an accurate description of the quantum mechanical system, if it is non-relativistic.  Then we expect that the universality of the cross section must persist even for the reltivistic case.  We show why the result should be generally valid, that the total absorption crossection must be proportional to the Planck area multiplied by a dimensionless pure number.    

Ultra-light particles as dark matter candidates (WISPs \cite{WISP}), have recently seen a resurgence.  With masses ranging from 10$^{-10}$ eV to 10$^{-21}$ eV, are routinely considered in the dark matter community.    WISPs would form quantum mechnical, gravitational bound states around black holes and neutron stars \cite{ABH,AD}.  Such quantum states of the WISPs would absorb or emit gravitons. 

\section{Quantum gravitons}
The perturbative quantum gravitational field corresponds to the tensor $h_{\mu\nu}$ which satisfies the linearized equation of motion
\beq
D_\sigma D^\sigma h_{\mu\nu}(x)=0\label{eom}
\eeq
where $D_\sigma$ is the covariant derivative in the Schwarzschild background.   The graviton field can be expanded in any complete set of orthonormal wave functions $\phi_n(x)$ that satisfy the equations of motion,
\beq
h_{\mu\nu}(x)=\sum_n \left(\phi_n(x) \epsilon_{\mu\nu}(n)A( n)+\phi_n^*(x)\epsilon^*_{\mu\nu}(n)A^\dagger( n)\right)
\eeq
where $\epsilon_{\mu\nu}(n)$ is the corresponding polarization tensor  and the operators $A^\dagger(n)$  and $A(n)$  satisfy the simple algebra of annihilation and creation operators
$
\left[A(n),A^\dagger(n')\right]=\delta_{nn'}
$.
These operators respectively create and destroy one  graviton in the state $\phi_n(x)$.

\section{Interaction Hamiltonian}
The interaction Hamiltonian is obtained following the anaylisis in \cite{gl} and \cite{bhgl} and we will not repeat it here in any detail. 
We consider the case of a very light WISP-like particle of mass $m$, bound to a much heavier massive particle of mass $M$ (this could even be a black-hole), so that the reduced mass $\mu=\frac{mM}{m+M}\approx m$.
In the perturbative, weak field approximation, the metric is
\beq
d\tau^2=\left(1-\frac{2GM}{r}\right)dt^2-\left(1+\frac{2GM}{r}\right)(\vec x\cdot \vec dx)^2/r^2-r^2d\Omega^2+h_{\m\n}dx^\m dx^\n.
\eeq
The classical dynamics of the light particle are then governed by the geodesic equation
\beq
\frac{d^2x^i}{d\tau^2}+\Gamma^i_{\mu\nu}\frac{dx^\mu}{d\tau}\frac{dx^\nu}{d\tau}=0
\eeq
where $x^\m$ are the coordinates of the light particle, and canonical quantization will generate the quantum theory, \cite{gl,bhgl,bertschinger}.    The Hamiltonian for the system is given by
\beq
H(x^i,\pi_j)=(1+\phi)E(p_j)\,\,{\rm where}\,\, E(p_j)=(\delta^{ij}p_ip_j+m^2)^{1/2}
\eeq
where $m$ is the WISP mass, and with $p_i=(1+\psi)\pi_i-(\delta^{ij}\pi_i\pi_j+m^2)^{1/2}w_i-s^j_{\,\,\, i}\pi_j$ which in our case becomes just $p_i=(1+\psi)\pi_i-s^j_{\,\,\, i}\pi_j$, see \cite{gl,bhgl} for the details.  Then we find our basic Hamiltonian of the WISP interacting with the gravitational field of the black hole
\beq
H_0(x^i,\pi_j)=\frac{|\pi|^2}{2m }+m \phi\label{hfree}
\eeq
and the interaction with the gravitational wave
\beq
H^\approx(x^i,\pi_j)=\frac{-h^{ ij}\pi _i\pi_j}{2m }.\label{hwave}
\eeq
The Schrödinger equation resulting from Eqn. \eqref{hfree} is that of the Newtonian hydrogen atom
\beq
\left(\frac{-\hbar^2}{2m }\nabla^2-\frac{GMm}{r}\right)\psi(x)=E\psi(x).\label{schrodinger}
\eeq
The time dependent perturbation that is relevant, from Eqn.\eqref{hwave}  (replacing $h_{\m\n}\rightarrow \frac{\sqrt{G}}{c^2}h_{\m\n}$ which is the canonically normalized metric perturbation) is given by
\beq
H^\approx(x^i,\pi_j)=- \frac{\sqrt{G}}{c^2}\frac{h^{ij}\pi_i\pi_j}{m }.
\eeq

\section{Amplitude for spontaneous or stimulated emission and cross section}
The amplitude for the absorption or for the stimulated emission of a graviton in a plane wave mode $\vec k,\epsilon$, where $\vec k$ is the momentum vector and $\epsilon$ is the polarization of the graviton,  is proportional to the matrix element  $\langle \psi_{f};(N_{\vec k}+1, k^\mu,\epsilon)|H^\approx(x^i,\pi_j)|\psi_{i};(N_{\vec k}, k^\mu,\epsilon)\rangle$ where a state $|\psi;(N_{\vec k}, k^\mu,\epsilon)\rangle$ corresponds to the light particle in a Newtonian hydrogen atomic state $\psi$ and $N_{\vec k}$ gravitons in a plane wave state $ k^\mu,\epsilon$ (there could be  many other occupied states in the system, that are just spectators) where $k^\mu$ is on shell, \ie $c|\vec k |=\omega$ .   Then using Fermi's golden rule we get the rate
\beq
\Gamma_{\phi_i\to\phi_n}=\frac{2\pi}{\hbar}\delta\lb E_f-E_i-\hbar\omega\rb |\langle \phi_{f};(N_{\vec k}+1, k^\mu,\epsilon)|H^\approx(x^i,\pi_j)|\phi_{i};(N_{\vec k}, k^\mu,\epsilon)\rangle|^2.
\eeq 
Modifying a calculation in Baym, \cite{gordon1990lectures} and following closely the calculation done in \cite{gl,bhgl} it is straightforward to obtain the total absorption cross section.  We have, assuming the graviton is + polarized and moving in the $z$ direction,
\beq
\Gamma_{\phi_i\to\phi_n}^{\rm abs}=\frac{2}{V}\sum_{\vec k}\frac{2\pi}{\hbar} \left(\frac{2 \pi\hbar c^2N_{\vec k}}{\omega}\right)\left(\frac{G}{c^4}\right)\left|\langle \phi_{f}|\left(\frac{\pi_x^2-\pi_y^2}{m}\right)|\phi_{i}\rangle\right|^2\delta\lb E_f-E_i-\hbar\omega\rb
\eeq
where $V$ is the volume of a large box in which everything happens and the factor of 2 in front is from summing over the contribution of the two polarizations of the gravitons.  Assuming a narrow beam subtending a solid angle $d\Omega$, and converting the sum over into an integral over $\omega$ (which is actually not done because of the delta function in energy) as $\frac{1}{V}\sum_{\vec k}\to d\Omega\int \frac{\omega^2d\omega}{(2\pi c)^3}$, we get
\beq
\Gamma_{\phi_i\to\phi_n}^{\rm abs}=d\Omega \frac{4\pi}{\hbar^2}\left(\frac{\omega^2}{(2\pi c)^3}\right) \left(\frac{2 \pi\hbar c^2N_{\vec k}}{\omega}\right)\left(\frac{G}{c^4}\right)\left|\langle \phi_{f}|\left(\frac{\pi_x^2-\pi_y^2}{m}\right)|\phi_{i}\rangle\right|^2
\eeq
where $\omega=\frac{E_f-E_i}{\hbar}$ and the delta function produces an additional factor of $\hbar$ in the denominator.  To get the total cross section we must divide by the total incoming flux of gravitons.  This corresponds to the intensity flux, the energy per unit area per unit frequency in the incoming beam  divided by the energy per graviton $\hbar\omega$, integrated over frequency, but as the beam is taken monochromatic, we effectively divide by the energy per graviton per unit frequency, $\hbar$
\beq
\Phi=\int d\omega \frac{I(\omega)}{\hbar\omega}= d\Omega\frac{\omega^4}{(2\pi c)^4}\frac{1}{\hbar}\left(\frac{2 \pi\hbar c^2N_{\vec k}}{\omega}\right)
\eeq
thus
\beq
\sigma=\frac{\Gamma_{\phi_i\to\phi_n}^{\rm abs}}{\Phi}=\frac{4\pi}{\hbar}\frac{2\pi c}{\omega^2}\left(\frac{G}{c^4}\right)\left|\langle \phi_{f}|\left(\frac{\pi_x^2-\pi_y^2}{m}\right)|\phi_{i}\rangle\right|^2\label{sigma}
\eeq
Now we can see the origin of the reason that the total absorbtion cross section is universal.  The matrix element 
\beq
\left|\langle \phi_{f}|\left(\frac{\pi_x^2-\pi_y^2}{m}\right)|\phi_{i}\rangle\right|
\eeq
must be proportional to the difference of energies involved in the transition, even including necessary intermediate states.  The  hydrogenic system
\beq
\left( -\frac{\hbar^2}{2m}\nabla_{\vec r}^2-\frac{GMm}{r}\right)\psi=E\psi
\eeq
can be scaled out to give
\beq
mc^2\left(-\frac{1}{2}\nabla_{\vec y}^2 -\frac{\bar\alpha}{y}\right)\psi=E\psi
\eeq
with $\vec y=\frac{mc}{\hbar}\vec r$ and $\bar\alpha=\frac{GMm}{\hbar c}$ is the effective fine structure constant.  Comparing with normal Coulombic hydrogen atom, we simply have $\bar\alpha\to \alpha=\frac{e^2}{\hbar c}\approx\frac{1}{137}$ and the energy levels are for the Coulomb problem and hence for the Newtonian problem
\beq
E_n=-\frac{mc^2\alpha^2}{2}\frac{1}{n^2}\to -\frac{mc^2\bar\alpha^2}{2}\frac{1}{n^2}
\eeq
where $n$ is the principal quantum number.   As all energies are proportional to $mc^2\bar\alpha^2$, they will neatly cancel out between the factor of $\omega^2$ in the denominator and the matrix element $\left|\langle \phi_{f}|\left(\frac{\pi_x^2-\pi_y^2}{m}\right)|\phi_{i}\rangle\right|^2$ in the numerator of Eqn.\eqref{sigma}.  This analysis is valid for the case of ultra-cold neutrons bouncing quantum mechanically on a level surface \cite{gl}, we will see explicitly that it is true for the case of the Newtonian hydrogen atom, and by extension it must be valid for the case of a quantum mechanical pendulum that is hanging in a gravitational field.  This last case is potentially experimentally significant, the mirrors in LIGO \cite{ligo} are found to be in a superposition very close to the  quantum mechanical ground state of the quantum mechanical pendulum that they represent \cite{ligomirrors}.  

\section{Calculation of the matrix element}
The matrix element to compute, for a + polarized and moving in the $z$ direction is given by
\beq
\langle \psi_{n',l',m'}|(\pi_x^2-\pi_y^2)|\psi_{n,l,m}\rangle
\eeq
where $\psi_{n,l,m}$ is the gravitational hydrogen atom wave function.  These are tabulated in any quantum mechanics textbook, but we record them here in detail for the sake of completeness:
\beq
\psi_{n,l,m}= Y_l^m(\theta,\varphi)R_{nl}(r)
\eeq
where 
\beq
R_{nl}(r)=N_{nl}\rho^lL^{2l+1}_{n-l+1}e^{-\rho/2}
\eeq
with $N_{nl}=\sqrt{(2/n a_0^*)^3(n-l-1)!/(2n(n+1)!)}$, $\rho=2r/na_0^*$ and the effective Bohr radius $a_0^*=\hbar^2/(GMm^2)$.  

The computation is considerably simplified using the identity, the idea found in Baym \cite{gordon1990lectures},
\beq
\left[ H,x\right]=\left[ \pi_x^2/2m,x\right]=(-i\hbar/m) \pi_x
\eeq
hence $\pi_x^2=-(m/\hbar)^2\left[ H,x\right]^2$ and corresponding for $y$.  Then the matrix element we need to calculate is given by (the energies only depend on the principal quantum number $n$)
\bea
&~&\langle \psi_{n',l',m'}|(\pi_x^2-\pi_y^2)|\psi_{n,l,m}\rangle=\langle \psi_{n',l',m'}|\left[ H,y\right]^2|\psi_{n,l,m}\rangle-\langle \psi_{n',l',m'}|\left[ H,x\right]^2|\psi_{n,l,m}\rangle\nn
&=&\sum_{n'',l'',m''} \left\{\la\psi_{n',l',m'}|\left[ H,y\right]|\psi_{n'',l'',m''}\rangle \la\psi_{n'',l'',m''}|\left[ H,y\right]|\psi_{n,l,m}\rangle\right.\nn
&-&\left.\langle \psi_{n',l',m'}|\left[ H,x\right]|\psi_{n'',l'',m''}\rangle\langle \psi_{n'',l'',m''}|\left[ H,x\right]|\psi_{n,l,m}\rangle\right\}\nn
&=&\sum_{n'',l'',m''} \left\{ (E_{n'}-E_{n''})(E_{n''}-E_{n})\la\psi_{n',l',m'}|y|\psi_{n'',l'',m''}\rangle \la\psi_{n'',l'',m''}|y|\psi_{n,l,m}\rangle\right.\nn
&-& \left. (E_{n'}-E_{n''})(E_{n''}-E_{n})\la\psi_{n',l',m'}|x|\psi_{n'',l'',m''}\rangle \la\psi_{n'',l'',m''}|x|\psi_{n,l,m}\rangle\right\}.\label{23}
\eea
Now the matrix elements of $x$ or $y$ between hydrogenic wave functions is rather simple, indeed,
$x=r\sin\theta\cos\varphi$ and $y=r\sin\theta\sin\varphi$ hence we can use the recurrence relation (found in Arfken, Eqn. 15.151 \cite{ARFKEN15}) 
\bea
e^{\pm i\varphi}\sin\theta Y^m_l(\theta,\varphi)=\mp\sqrt{\frac{(l\pm m+1)(l\pm m+2)}{(2l+1)(2l+3)}}Y^{m\pm1}_{l+1}(\theta,\varphi) \pm\sqrt{\frac{(l\mp m)(l\mp m-1)}{(2l-1)(2l+1)}}Y^{m\pm 1}_{l-1}(\theta,\varphi) \nonumber\\
\eea
which yields for the matrix element
\bea
\la\psi_{n'',l'',m''}|re^{\pm i\varphi}\sin\theta|\psi_{n,l,m}\rangle\nn
\ame \mp\sqrt{\frac{(l\pm m+1)(l\pm m+2)}{(2l+1)(2l+3)}}\delta_{m'',m\pm 1}\delta_{l'',l+ 1}\int dr r^2 R_{n'',l+1}rR_{n,l}\nn
&\pm& \sqrt{\frac{(l\mp m-1)(l\mp m)}{(2l+1)(2l-1)}}\delta_{m'',m\pm 1}\delta_{l'',l-1}\int dr r^2 R_{n'',l-1}rR_{n,l}
\eea
we note:
\begin{equation*}
    \hspace{-2cm}\langle \psi_{n',l',m'}|y|\psi_{n'',l'',m''}\rangle\langle \psi_{n'',l'',m''}|y|\psi_{n,l,m}\rangle-\langle \psi_{n',l',m'}|x|\psi_{n'',l'',m''}\rangle\langle \psi_{n'',l'',m''}|x|\psi_{n,l,m}\rangle=
\end{equation*}
\begin{equation*}
    \hspace{-3cm}-\frac{1}{2}[\langle \psi_{n',l',m'}|re^{i\varphi}\sin\theta|\psi_{n'',l'',m''}\rangle\langle \psi_{n'',l'',m''}|re^{i\varphi}\sin\theta|\psi_{n,l,m}\rangle+
\end{equation*}
\begin{equation}
    \hspace{3cm}+\langle \psi_{n',l',m'}|re^{-i\varphi}\sin\theta|\psi_{n'',l'',m''}\rangle\langle \psi_{n'',l'',m''}|re^{-i\varphi}\sin\theta|\psi_{n,l,m}\rangle]
\end{equation}
The general result is a sum over $n'',l'',m''$ in Eqn. \eqref{23}.   For simplicity, we present it for a transition from the ground state with $n,l,m=1,0,0$, we obtain:
\begin{equation}
    \langle\psi_{n',l',m'}|(\pi_x^2-\pi^2_y)|\psi_{1,0,0}\rangle=
\end{equation}
\begin{equation*}       
    =\left(\frac{m}{\hbar}\right)^2\sqrt{\frac{2}{15}}\delta_{l',2}(\delta_{m',2}+\delta_{m',-2})\sum_{n''=2}^\infty(E_{n''}-E_{n'})(E_1-E_{n''})\left(\int r^2drR_{n',2}rR_{n'',1}\right)\left(\int r^2drR_{n'',1}rR_{1,0}\right)    .
\end{equation*}
Hence
\bea
    &~&\langle\psi_{n',l',m'}|\left(\frac{\pi_x^2-\pi^2_y}{m}\right)|\psi_{1,0,0}\rangle=\nonumber\\
    \ame\frac{G^2M^2m^3}{4\hbar^2}\sqrt{\frac{2}{15}}\delta_{l',2}(\delta_{m',2}+\delta_{m',-2})\sum_{n''=2}^\infty\bigg(\frac{1}{n''^2}-\frac{1}{n'^2}\bigg)\bigg(1-\frac{1}{n''^2}\bigg)I_{n',n''}
\eea
where:
\bea
    I_{n',n''}\ame\frac{2^7}{(n'n''^2)^3}\frac{(n''-2)!}{(n''+1)!}\sqrt{\frac{(n'-3)!}{(n'+2)!}}\int_0^\infty du\,u^6e^{-u(n''^{-1}+n'^{-1})}L^5_{n'-3}\bigg(\frac{2}{n'}u\bigg)L^2_{n''-2}\bigg(\frac{2}{n''}u\bigg)\times\nonumber\\
 &\times&\int_0^\infty d\xi\,\xi^4e^{-\xi(1+n''^{-1})}L^{3}_{n''-2}\bigg(\frac{2}{n''}\xi\bigg)
\eea
To get a better feel of what this general expression implies, we do the explicit calculation for transitions between the ground state and the second excited state (note we need to have $\Delta l=2$ for a non-zero transiton amplitude.  Thus for $n'=3$ and $n=1$  we find:
\begin{equation}
    \int_0^\infty d\xi\,\xi^4e^{-\xi(1+n''^{-1})}L^{3}_{n''-2}\bigg(\frac{2}{n''}\xi\bigg)=2 (n''-1)^{( n''-2)} n''^7 (1 + n'')^{(-2 - n'')}
\end{equation}
\begin{equation}
    \int_0^\infty du\,u^6e^{-u(n''^{-1}+3^{-1})}L^2_{n''-2}\bigg(\frac{2}{n''}u\bigg)=26244\frac{n''^9(n''^2-1)(n''-3)^{(n''-4)}}{(n''+3)^{(n''+4)}}
\end{equation}
where these integrals are obtained by using the expression for the Laguerre polynomials found in Arfken Eqn. 18.59 \cite{ARFKEN18}, given as finite sums, and then integrating termwise.  The resulting expressions, also  finite sums,  can be summed analytically, we did however use Mathematica to perform the sums.  We find
\bea
   &~&\langle\psi_{3,l',m'}|\left(\frac{\pi_x^2-\pi^2_y}{m}\right)|\psi_{1,0,0}\rangle=\nonumber\\
  \ame\frac{G^2M^2m^3}{4\hbar^2}\delta_{l',2}(\delta_{m',2}+\delta_{m',-2})\sqrt{\frac{2}{15}}\frac{2^7\cdot5832}{3^{3}\cdot\sqrt{5!}}\sum_{n''=2}^\infty -n''^5\frac{(n''-3)^{n''-3}}{(n''+3)^{n''+3}}\frac{(n''-1)^{n''-1}}{(n''+1)^{n''+1}}\label{37}
\eea
where
\begin{equation}
    \sum_{n''=2}^\infty -n''^5\frac{(n''-3)^{n''-3}}{(n''+3)^{n''+3}}\frac{(n''-1)^{n''-1}}{(n''+1)^{n''+1}}\approx0.000276583351\label{38}
\end{equation}
Then using these calculations, for the total absorption cross section for a transition from the ground state, $n=1$, to the second excited state, $n=3$, (or equivalently the emission probabilty for the reverse transition) we find the result, from Eqn\eqref{sigma}, Eqn\eqref{37} and Eqn\eqref{38}, 
\beq
\sigma = 8\pi^2\frac{\hbar G}{c^3}\frac{\left(\frac{G^2M^2m^3}{4\hbar^2}\right)^2}{\left(\frac{mc^2}{2}\left(\frac{GMm}{\hbar c}\right)^2\left(1-\frac{1}{9}\right)\right)^2}\beta=\frac{\hbar G}{c^3}\beta
\eeq
where 
\beq
\beta=\frac{81\pi^2}{32}\left(\sqrt{\frac{2}{15}}\frac{2^7\cdot5832}{3^{3}\cdot\sqrt{5!}}\sum_{n''=2}^\infty -n''^5\frac{(n''-3)^{n''-3}}{(n''+3)^{n''+3}}\frac{(n''-1)^{n''-1}}{(n''+1)^{n''+1}}\right)^2\approx 1.91
\eeq 
is a pure number that is parametrically of order 1, that depends on the specific geometry of the quantum system.    

\section{Discussion}
For the case of  quantum bouncers  \cite{gl}, $\beta$ is a function of the difference of the zeros of the Airy function, for the case here with particles bound in the Schrodinger levels of essentially a Newtonian, hydrogen atom, $\beta$ is somewhat more complicated, being a function of a difference of the principal quantum numbers involved in the transition including the principal quantum numbers of intermediate states.   We imagine for the case of a quantum pendulum, the result will be of the same kind and rather simple.

In all cases, the total absorption cross section on a source that is gravitationally defined, will be universal in that it is independent of the masses of the particles involved.  The cross section is proportional to the Planck area $\frac{\hbar G}{c^3}$.    This universality is surely an expression of the equivalence principle now applied to quantum mechanical systems.  

We imagine that we can apply this result to the case of gravitons interacting  with a quantum mechanical detector as given for example by LIGO \cite{ligo}.  The Planck area is aproximately $\frac{\hbar G}{c^3}\approx 2.61\times 10^{-70}m^2$ thus actually seeing the shot noise from relic gravitons seems very difficult.  However, there have been analyses of detecting exactly this type of gravitational radiation \cite{grishchuk,beckwith,rg}.   The graviton shot noise will have a temperature of the order of $\approx {10^{-13}}^\circ$ K, assuming decoupling at the Planck scale, $10^{19}$ GeV, which although small, is not ridiculously small.  Nano-Kelvin temperatures are routinely measured in the lab.  The relic shot noise will leave a specific detection profile which could be analyzed through appropriate filtering and template comparisons.

\section{Acknowledgments} We thank Jorge Gamboa for discussions, who initially a participated in this work.  We thank the Scholarship Program for Emerging Leaders in the Americas of the Government of Canada for enabling BA-L's visit to the Université de Montréal. We also thank Affaires Internationales de l’Université de Montréal for facilitating the academic exchange with the Universidad de Santiago and NSERC Canada for their financial support. We thank, for their hospitality, Jorge Gamboa of the Physics Department of the Universidad de Santiago, Santiago, Chile, and Richard Easther of the Department of Physics of the University of Auckland, Auckland, New Zealand (where this work was written up).

\bibliographystyle{apsrev}
\bibliography{graviton-scatt-GRFEC2025-new-version}

\end{document}